\documentclass{article}
\usepackage[utf8]{inputenc}
\usepackage[english]{babel}
\usepackage[T1]{fontenc}
\usepackage{longtable}
\usepackage{amssymb}
\usepackage{amsthm}
\usepackage{enumerate}
\usepackage{amsmath}
\newtheorem{problem}{Problem}
\newtheorem{innercustomthm}{Theorem}
\newenvironment{customthm}[1]
  {\renewcommand\theinnercustomthm{#1}\innercustomthm}
  {\endinnercustomthm}
\newtheorem*{cor}{Corollary}
\usepackage{url}

\usepackage[a4paper,top=2.54cm,bottom=2.54cm,left=2.54cm,right=2.54cm]{geometry}
\usepackage{indentfirst} 
\usepackage[colorlinks=true, allcolors=blue]{hyperref}

\title{\textbf{On the Fairness of Name-Based Rationing System for Purchases of Masks Policy}}
\author{Yu-Ting Liu}
\date{February 5, 2020}

\begin{document}
\maketitle

\section{Introduction}
Owing to the worsening of 2019 Novel Coronavirus (2019-nCoV) outbreak in China, panic buying of face masks had shown an accelerated growth worldwide, especially in the Greater China region. Facing this tough situation of possible shortage of masks, the Central Epidemic Command Center (CECC) in Taiwan announced a name-based rationing system for masks on February 3, 2020, which would be launched three days after the announcement. The policy was expected to be capable of ensuring universal access to face masks as well as fairness and transparency of resources allocation and curbing the hoarding of face masks.

\section{Preliminary}
In the announcement, CECC pointed out several changes in the sale of masks, one of which is quoted as follows:
\begin{quote}\itshape
    ``To reduce the queue for face masks, the policy allows the people whose ID card number ends with an even number to buy masks on Tuesdays, Thursdays and Saturdays, and the people whose ID card number ends with an odd number to buy masks on Mondays, Wednesdays and Fridays while everyone can buy them on Sundays. Furthermore, office workers and people with disabilities can ask their family members or friends to buy face masks on their behalf. However, each person is allowed to buy face masks for another person by presenting the person’s IC card, and the abovementioned rules of purchase apply. Moreover, face masks for children are only allowed to be purchased when IC cards of children under 12 are presented.''
\end{quote}
In short, all the people would be partitioned into two disjoint sets, and
\begin{itemize}
    \item on Tuesdays, Thursdays and Saturdays, only people whose ID card number ends with an even number are allowed to buy masks;
    \item on Mondays, Wednesdays and Fridays, only people whose ID card number ends with an odd number are allowed to buy masks.
\end{itemize}
Notice that a valid ID number is a capital letter, which depends on the place of one's first household registration, followed by nine digits, where each digit ranges from 0 to 9, the first digit recognizes the gender and the last digit is a checksum depending on the leading letter and the other eight digits, one should be dubious about the ``fairness'' of this policy for the two disjoint sets of people considering the nature of checksum computations, i.e., the distribution of the parity of the last digit of ID number is non-trivial and remains unknown. 
\begin{table}[t]
    \centering
    \renewcommand{\arraystretch}{1.25}
    \begin{tabular}{c||c|c|c|c|c|c|c|c|c|c|c|c|c|c|c|c|c|c|c}
        $\mathcal{A}$ & A & B & C & D & E & F & G & H & I & J & K & L & M & N & O & P & Q & R & S\\\hline
        $\alpha$ & 10 & 11 & 12 & 13 & 14 & 15 & 16 & 17 & 34 & 18 & 19 & 20 & 21 & 22 & 35 & 23 & 24 & 25 & 26
    \end{tabular}\vspace{8pt}
    \begin{tabular}{c||c|c|c|c|c|c|c}
        $\mathcal{A}$ & T & U & V & W & X & Y & Z\\\hline
        $\alpha$ & 27 & 28 & 29 & 32 & 30 & 31 & 33
    \end{tabular}
    \caption{One-to-one correspondence between a capital letter and an integer.}
    \label{11}
\end{table}
\section{System Model}
Let $\mathcal{A}a_1a_2a_3a_4a_5a_6a_7a_8a_9$ be a valid ID number, where $\mathcal{A}$ is a capital letter and has a one-to-one correspondence to an integer $10\leq\alpha\leq35$, shown in Table \ref{11}, and $a_i$ is an integer in the range 0 to 9 for each $i=1,2,\cdots,9$. Let $0\leq \alpha'\leq9$ satisfies $\alpha\equiv\alpha'\ \ (\text{mod }10)$, then
\begin{align}\label{rule10}
    \hspace*{1cm}a_9&\equiv\frac{\alpha-\alpha'}{10}+9\alpha'+\sum^8_{i=1}(9-i)a_i&&\hspace*{-1cm}(\text{mod }10)\nonumber\\[.25em]
    \hspace*{1cm}&\equiv\frac{\alpha-11\alpha'}{10}+\sum^8_{i=1}(9-i)a_i&&\hspace*{-1cm}(\text{mod }10).\hspace*{2cm}
\end{align}
Using \eqref{rule10}, the fairness issue can be reformulated as follows:
\begin{problem}\label{P1}
Let $f_X(x)$ be the probability mass function of the parity of $a_9$, then
\begin{equation*}%\label{density}
    f_X(x)=\left\{\begin{array}{ll}
        p, & \text{if }a_9\equiv0\ \ (\text{mod }2) \\[.25em]
        1-p, & \text{if }a_9\equiv1\ \ (\text{mod }2),
    \end{array}\right.
\end{equation*}
where $x=a_9\ \ \text{mod }2$, for some $0\leq p\leq1$. If $p$ is far from $1/2$, then the policy could be considered unfair, since, for example, if $p$ were equal to 0.9, then $90\%$ of the people could go to the pharmacies or drugstores to buy masks on Mondays, which would lead to a severe competing purchase. So, is $p$ close enough to one half or not in reality?
\end{problem}

To be more precise, since $p$ is the probability that $a_9$ is even, also, 10 is a multiple of 2, let 
\begin{equation*}
    b_i=\left\{\begin{array}{ll}
        0, & \text{if $a_i$ is even} \\
        1, & \text{if $a_i$ is odd},
    \end{array}\right.
\end{equation*}
for $1\leq i\leq9$, then \eqref{rule10} implies
\begin{align}\label{rule2}
    \hspace*{1cm}b_9&\equiv\frac{\alpha-11\alpha'}{10}+\sum^8_{i=1}(9-i)b_i&&\hspace*{-1cm}(\text{mod }2)\nonumber\\[.25em]
    \hspace*{1cm}&\equiv\frac{\alpha-11\alpha'}{10}+b_2+b_4+b_6+b_8&&\hspace*{-1cm}(\text{mod }2).\hspace*{2cm}
\end{align}
From \eqref{rule2} we know that the parity of the last digit of ID number only depends on the leading capital letter as well as the second, fourth, sixth and eighth digits. To briefly discuss the fairness issue, we may assume that, in a specific region, the ID numbers of most of the people start with the same capital letter, then $p$ only depends on the parity of $b_2+b_4+b_6+b_8$, and we get an equivalent problem as follows:
\begin{problem}\label{P2}
Without loss of generality, assume in Hsinchu City, most of the people have ID number starting with ``O'', from Table \ref{11} we know that $\alpha=35$ and $\alpha'=5$. Since $\frac{35-11\times5}{10}=-2$ is an even number, $b_9=b_2+b_4+b_6+b_8\ \ \text{mod }2$. Let $p_j$ denotes the probability that $a_j$ is even, for $j=2,4,6,8$, and assume that $a_j$'s are independent, then
\begin{align*}
    p&=p_2p_4p_6p_8+p_2p_4(1-p_6)(1-p_8)+p_2(1-p_4)p_6(1-p_8)+p_2(1-p_4)(1-p_6)p_8\\
    &\quad+(1-p_2)p_4p_6(1-p_8)+(1-p_2)p_4(1-p_6)p_8+(1-p_2)(1-p_4)p_6p_8\\
    &\quad+(1-p_2)(1-p_4)(1-p_6)(1-p_8).
\end{align*}
What is condition for $p_2,p_4,p_6,p_8$ such that $p$ is close to or exactly one half?
\end{problem}
\section{Fairness Condition}
One may start from an easy case. Let $y,x_1,x_2\in\mathbb{Z}_2$, where $\mathbb{Z}_n$ denotes the group of integers modulo $n$ for some positive integer $n$. If $x_1,x_2$ are independent variables and $y=x_1+x_2$, then
\begin{align*}
    \mathbb{P}\left\{y=0\right\}&=\mathbb{P}\{x_1=0\}\mathbb{P}\{x_2=0\}+\mathbb{P}\{x_1=1\}\mathbb{P}\{x_2=1\}\\[.25em]
    &=\mathbb{P}\{x_1=0\}\mathbb{P}\{x_2=0\}+\Big(1-\mathbb{P}\{x_1=0\}\Big)\Big(1-\mathbb{P}\{x_2=0\}\Big)\\[.25em]
    &=2\left(\mathbb{P}\{x_1=0\}-\frac{1}{2}\right)\left(\mathbb{P}\{x_2=0\}-\frac{1}{2}\right)+\frac{1}{2}
\end{align*}
It is obvious that if $x_1$ {or} $x_2$ is 0 of probability 1/2, then $\mathbb{P}\big\{y=0\big\}$ is always 1/2.
\begin{customthm}{1}\label{thm}
Consider a sequence of independent variables $x_1,x_2,\cdots,x_m\in\mathbb{Z}_2$, for some $m\geq2$, let $y^{(f)}\in\mathbb{Z}_2$ defined as $y^{(f)}=\sum\limits^f_{k=1}x_k$ for some $2\leq f\leq m$ and $q_k=\mathbb{P}\{x_k=0\}$ for $1\leq k\leq m$. Then
\begin{equation}\label{thm1}
    \mathbb{P}\left\{y^{(f)}=0\right\}=2^{f-1}\prod^f_{k=1}\left(q_k-\frac{1}{2}\right)+\frac{1}{2}.
\end{equation}
\end{customthm}\noindent
\textit{Proof.} \begin{enumerate}[(i)]
    \item It is clear that \eqref{thm1} holds for $f=2$.
    \item If \eqref{thm1} still hold for some $2\leq f=t<m$, consider $f=t+1$, then $y^{(t+1)}=y^{(t)}+x_{t+1}$ and
    \begin{align*}
        \mathbb{P}\left\{y^{(t+1)}=0\right\}&=\mathbb{P}\left\{y^{(t)}=0\right\}\cdot q_{t+1}+\left(1-\mathbb{P}\left\{y^{(t)}=0\right\}\right)\cdot\big(1-q_{t+1}\big)\\[.5em]
        &=1-q_{t+1}-\mathbb{P}\left\{y^{(t)}=0\right\}+2\mathbb{P}\left\{y^{(t)}=0\right\}\cdot q_{t+1}\\[.5em]
        &=2\mathbb{P}\left\{y^{(t)}=0\right\}\left(q_{t+1}-\frac{1}{2}\right)-q_{t+1}+1\\[.5em]
        &=2^t\prod^{t+1}_{k=1}\left(q_k-\frac{1}{2}\right)+q_{t+1}-\frac{1}{2}-q_{t+1}+1\\[.5em]
        &=2^t\prod^{t+1}_{k=1}\left(q_k-\frac{1}{2}\right)+\frac{1}{2}.
    \end{align*}
    \item By induction, \eqref{thm1} holds for all $2\leq f\leq m$.
\end{enumerate}\qed
\begin{cor}
Consider the notations defined in Theorem \ref{thm}. Then, for $2\leq f\leq m$, $\mathbb{P}\left\{y^{(f)}=0\right\}\neq\frac{1}{2}$ if and only if $q_k\neq\frac{1}{2}$ for all $k\leq f$.
\end{cor}
This corollary, which is a trivial result following from \eqref{thm1}, implies that, in Problem \ref{P2}, if none of $p_2,p_4,p_6,p_8$ is $\frac{1}{2}$, then $p\neq\frac{1}{2}$. Finally, we can conclude that, if at least one of the distributions of parities of the second, fourth, sixth or eighth digit of ID number is uniform (i.e., odd and even are equally-likely), then the policy could be considered fair.

A special case can also be found in \cite{gallager}, Gallager showed that if a sequence of $m$ independent binary digits in which every digit is a 1 with probability $q$, then the probability that an even number of digits are 1 is
\begin{equation*}
    \frac{1+(1-2q)^m}{2}.
\end{equation*}
That is, even if $q=1/3$, when $m=4$, the probability is 0.5062 which is only slightly greater than 1/2.

\section{Conclusion}
If one can show the evidence of one of the exact distributions of parities of the second, fourth, sixth and eighth digits, the fairness issue of this policy can be confirmed. Nevertheless, as mentioned and calculated in the previous paragraph, people should always try not to worry and be frustrated about it given that the two disjoint sets of people are probably of the same size.

Despite the fact that the above derivations are strict enough, mathematically, there may still exist some exceptions. Note that in reality, the second digits are mostly in the range 0 to 2; also, due to the entendre between numbers and Chinese words, the number ``4'' may occur less than other numbers in the ID number.

\bibliographystyle{unsrt}
\nocite{news,ID}
\bibliography{gallager,news,ID}

\begin{thebibliography}{1}

\bibitem{gallager}
Robert Gallager.
\newblock Low-density parity-check codes.
\newblock {\em IRE Transactions on information theory}, 8(1):21--28, 1962.

\bibitem{news}
\url{https://www.mohw.gov.tw/cp-4638-51319-2.html}.

\bibitem{ID}
\url{https://www.moi.gov.tw/chi/chi_faq/faq_detail.aspx?t=2&n=14779&p=6&f=2}.

\end{thebibliography}
\end{document}